# Stabilization and Destabilization of Multimode Solitons in Nonlinear Degenerate Multi-Pass Cavities

Junhan Huang[1,2], Bingbing Zhu[1,2], Shanyue Li[1,2], Kun Ding[1,2], and Zhensheng Tao[1,2*]

[1] State Key Laboratory of Surface Physics, Key Laboratory of Micro and Nano Photonic Structures (MOE), and Department of Physics, Fudan University, Shanghai 200433, China.

[2] Shanghai Key Laboratory of Metasurfaces for Light Manipulation, Fudan University, Shanghai 200433, China.

*Corresponding author. Email: zhenshengtao@fudan.edu.cn.

**Abstract**

Optical solitons in multimode nonlinear optical systems offer a unique platform for exploring the interplay of nonlinearity, dispersion, and spatial mode coupling, offering insights into complex nonlinear wave phenomena. Multi-pass cavities (MPCs) incorporating nonlinear Kerr media serve as prototypical systems, enabling high-efficiency supercontinuum generation and pulse compression. However, stabilizing femtosecond laser pulses in solid-medium-based MPCs (solid MPCs) under strong Kerr nonlinearity remains a significant challenge due to multimode coupling, which disrupts beam stability. In this work, we address this challenge by investigating the stability of laser pulses in MPCs using Floquet and perturbation model. We identify novel mode-coupling-suppression (MCS) medium lengths, where destructive interference among multimode wave components suppresses coupling and facilitates soliton stabilization. Under MCS conditions, our simulations demonstrate stable beam propagation in solid MPCs with nonlinear phases up to $1.5\pi$ per pass, achieving >13-fold pulse compression with excellent spatio-spectral homogeneity. Our findings offer valuable guidance for designing advanced MPCs with tailored Kerr media.

**Introduction**

Optical solitons – self-sustained, localized wave-packets stabilized by a balance between dispersion and nonlinearity – have been extensively studied due to their crucial role in high-speed optical communication[1], ultrafast pulse generation[2,3], and other nonlinear optical phenomena[4]. Of particular interest are multimode solitons, which involve interactions among multiple spatial modes, and exhibit rich spatiotemporal dynamics such as spatial self-organization[5–7], soliton fission[8,9], and the emergence of spatiotemporal solitons, also known as "light bullets"[10–13].

While multimode solitons have been extensively studied in optical fibers[14–16], multi-pass cavities (MPCs) offer a complementary platform with weak transverse mode confinement and exceptional potential for high-energy applications[17,18] (Fig. 1a). By circulating femtosecond laser pulses, nonlinear MPCs enable substantial spectral broadening, leading to high-throughput, and high-quality pulse compression with high power[19,20], high pulse energy[20], and few-cycle durations[21,22]. Commonly used Kerr media include noble gases[19–21,23–42], and single[22,43–53] or multiple[54–58] solid media. Furthermore, MPCs allow the access to diverse nonlinear regimes by precisely controlling dispersion through cavity mirrors or varying the nonlinear optical media.

A central design challenge of nonlinear MPCs is the long optical path length required to sustain the large number of roundtrips inside the cavity (typically 30~50). This requirement arises from the intrinsically low single-pass nonlinear phase (SNLP) required to maintain high spatio-spectral quality after nonlinear spectral broadening. The limitation is particularly evident in solid-medium-based MPCs (solid MPCs), where strong nonlinearity readily induces spatio-spectral degradation, limiting operation to SNLP values below ~0.8π. In contrast, gas-filled MPCs can tolerate much higher SNLPs – up to ~2π (Fig. 1b). The physical origin of this striking performance gap between solid and gas-filled MPCs, however, remains poorly understood. Fundamentally, resolving this question is equivalent to identifying stable soliton solutions in MPCs operating under strong nonlinearity.

Previous theoretical studies have highlighted the role of multimode coupling in spatial degradation, both in single-pass setups[59] and gas-filled MPCs[60,61]. Coupled-mode theory has demonstrated that degenerate cavity geometries – where higher-order modes are phase-matched with the fundamental modes – strongly influence beam stability[61,62]. This insight is particularly important, as most nonlinear MPC implementations utilize degenerate cavity configurations, also known as "$q$-preserving" geometries[63,64]. In addition, soliton-like dynamics in pulse propagation have been observed in solid MPCs at mid-infrared wavelengths[43]. Despite these insights, the underlying mechanisms that stabilize or destabilize multimode solitons remain elusive.

In this work, we theoretically investigate laser-beam stability in nonlinear MPCs with strong Kerr nonlinearity from the perspective of multimode solitons. Using Floquet analysis and first-order perturbation theory, we elucidate the mechanisms by which multimode coupling, induced by the interplay between cavity degeneracy and Kerr nonlinearity, destabilizes solitons (Fig. 1c). These analytical insights are supported by full numerical simulations that incorporate space-time coupling effects. Our results clarify the observed differences between gas-filled and solid MPCs. More importantly, we identify a novel concept of *mode-coupling-suppression (MCS) medium lengths* ($d_{MCS}$), where destructive interference among multimode wave components within the Kerr medium suppresses multimode coupling. This stabilization mechanism facilitates soliton formation in degenerate nonlinear MPCs under high nonlinearity (Fig. 1d), with gas-filled MPCs emerging as a special case of the MCS condition.

Under the MCS condition, we demonstrate stable soliton propagation with a nonlinear phase up to $b=1.50\ \pi$ per pass in solid MPCs (star in Fig. 1b), resulting in >13-fold pulse compression with excellent spatio-spectral quality requiring only 9 roundtrips. This result exceeds the SNLP limits of the existing solid MPCs, offering valuable insights for designing advanced nonlinear MPCs with tailored Kerr media.

**Phase Diagrams of Beam-Propagation Stability**

The conceptual schematic of a nonlinear MPC is shown in Fig. 1a. It consists of two concave mirrors (C.M.), each with a focal length $F$, separated by a distance of $2L$. At the cavity center, a Kerr medium with a length of $2d$ interacts with femtosecond laser pulses, inducing an SNLP denoted as $b$. Here, the nonlinear phase is characterized by $b = \frac{2\pi}{\lambda_0} n_2 \int_{-d}^{d} I(z)dz$, where $\lambda_0$ is the laser wavelength, $n_2$ is the nonlinear refractive index, $z$ is the coordinate along the cavity axis (with the origin at the cavity center), and $I(z)$ is the laser intensity within the Kerr medium.

To investigate laser-beam propagation stability, we perform full space-time-coupled nonlinear Schrödinger equation (NLSE) simulations[65] over a range of conditions (see Methods). The accuracy of the simulations is benchmarked against the spectral measurements and transform-limited pulse durations reported in previous experimental studies (see SM Section S1). In the cases shown in Figs. 2a-c, the cavity length is fixed at 2$L$=79.5 cm, while the focal length $F$ and medium length 2$d$ are varied. When $d/L \to 0$ by setting 2$d$=1 mm (Fig. 2a), the simulation represents typical conditions of solid MPCs, where the medium length is much smaller than the cavity length. In contrast, $d/L$=1.0 (Fig. 2c) corresponds to a gas-filled MPC, where the gas medium occupies the entire cavity volume.

The simulations use $\tau_p$=170 fs pulses at a center-wavelength of $\lambda_0$=1030 nm, representative of a typical Yb:KGW femtosecond laser. For each cavity geometry, the input beam profile is set to the fundamental Laguerre-Gaussian (LG) mode of the corresponding linear cavity, with a beam waist given by $w_0 = \frac{\sqrt{\lambda_0 L}}{\pi} \sqrt{\frac{2F}{L} - 1}$. To reveal the fundamental physics of beam propagation in nonlinear MPCs, group-delay dispersion (GDD) of the Kerr medium is neglected in these simulations. The results including realistic material GDD are presented later.

In the simulations, the SNLP $b$ is adjusted by varying the input pulse energy $E_0$ according to

$$b = \frac{8 n_0 n_2 d_{\text{eff}}}{w_0^2 \lambda_0} \frac{E_0}{\tau_p}, \tag{1}$$

where $n_0$ is the refractive index of the Kerr medium, and $d_{\text{eff}} = z_0 \arctan\left(\frac{d}{z_0}\right)$ represents an effective medium length[63], with $z_0$ being the Rayleigh length. Details of the NLSE simulations are provided in Methods.

For the simulations using thin Kerr media (Fig. 2a), achieving large $b$ values requires laser peak powers ($P_0 = E_0/\tau_p$) exceeding the material critical power $P_{\text{cr}}$, reaching up to $P_0/P_{\text{cr}} \approx 15$. Although such powers can induce self-focusing, the short medium length ensures the focal point lies outside the medium, preventing beam collapse and material damage in practice[66–68]. For thicker media (Figs. 2b-c), the peak power is typically below 50% of the critical power, which is sufficient to achieve the targeted *b* values.

We quantitatively assess beam instability by evaluating the spatio-spectral homogeneity ($\langle V \rangle$) of the output beam[57] (see Methods). Figures 2a-c present the two-dimensional plots of output-beam inhomogeneity (1-$\langle V \rangle$) after 10 roundtrips as functions of *b* and *F*/*L* for different *d*/*L* values.

Several key observations can be made. First, the spatio-spectral homogeneity improves as *d*/*L* increases. Second, for $d/L \to 0$ (Fig. 2a), cavities with higher degeneracy exhibit pronounced spatio-spectral inhomogeneity. Here, cavity degeneracy arises when the accumulated Gouy phases of the *g*- and *n*-th order LG modes after a single pass satisfy $\xi_n = \xi_g + 2j\pi$, where *j* is an integer. Since Gaussian-profiled laser beams are usually used as inputs, the degree of degeneracy relative to the $LG_{00}$ mode is of particular interest. The Gouy-phase condition $\xi_n = \xi_0 + 2j\pi$ leads to the following degeneracy condition:

$$4u \arctan\left(\frac{1}{\sqrt{2F/L-1}}\right) = 2v\pi, \qquad (2)$$

where (*u*, *v*) are a pair of coprime integers (*v*<*u*), representing the cavity-degeneracy indices. For a Herriott-type MPC[69] in practice, the *q*-preserving configuration results in closed ray paths, where rays retrace their trajectories after a certain number of roundtrips[62,63]. Under this condition, the degeneracy index *u* represents the total number of laser spots present on each cavity mirror (see inset of Fig. 1a). The index *v* indicates the position where the ray bounces after its first roundtrip (blue spot), assuming the incident beam initially hits spot #1 (green spot).

Quantitatively, the degree of cavity degeneracy can be characterized by a normalized density of states (DOS), which counts the number of LG modes that are degenerate with the $LG_{00}$ mode (see Methods). A clear correlation is observed between the DOS peaks (Fig. 2d) and spatio-spectral inhomogeneity patterns (Fig. 2a). This observation is further corroborated by analyzing the spatio-temporal profiles for both degenerate (condition A) and non-degenerate (condition B) cases (Figs. 2e-f). Under the degenerate conditions, higher-order modes and pulse splitting are evident (Fig. 2e), indicating spatio-temporal instability.

Lastly, but more interestingly, we find, at specific medium lengths, denoted as $2d_{MCS}$, beam quality can be significantly improved even under the degenerate conditions. For instance, when $d_{MCS} = \frac{1}{3}L$ with $F/L=2/3$ and degeneracy indices ($u$, $v$)=(3, 2) (condition C in Fig. 2b), the strong inhomogeneity observed in condition A (Fig. 2a) is remarkably suppressed, resulting in a spatio-spectrally homogeneous output beam. The comparison between Figs. 2g and 2e further demonstrates that the spatio-temporal breakdown is mitigated under $d_{MCS}$.

To highlight this intriguing behavior, we further compare the zoomed-in views around the degenerate point ($F/L=2/3$) for both $d/L \to 0$ and $2d=2d_{MCS}$ (Figs. 2h-i). Clearly, at $2d=2d_{MCS}$, the inhomogeneity at the degenerate point is sharply suppressed, while the surrounding regions with lower DOS exhibit some inhomogeneity, forming a "V"-shape pattern around the degenerate point (Fig. 2i). Notably, the high-homogeneity regions in Figs. 2a-c indicate the formation of discrete spatial solitons, where laser beams maintain stable spatial profiles within the nonlinear medium or on the cavity mirrors (see SM Section S2).

**Floquet and Perturbation Model Analysis**

While our numerical simulations offer valuable insights into the stability landscape of discrete spatial solitons in nonlinear MPCs, the underlying mechanisms remain intricate. To elucidate these mechanisms, we employ analytical approaches based on Floquet and perturbation theories, which neglect space-time coupling effects. Despite this simplification, we will show that these frameworks can accurately capture the key stability criteria and provide a robust foundation for interpreting soliton behavior in nonlinear MPCs.

Given that an optical cavity represents the periodic propagation of a light beam in space, its behavior can be described using Floquet theory[70]. In the linear regime, the system is governed by the Floquet eigenequation:

$$\left[H_0(r,z) - i\frac{\partial}{\partial z}\right]|\Phi_{n,m}(r,z)\rangle = \varepsilon_{n,m}|\Phi_{n,m}(r,z)\rangle, \qquad (3)$$

where $H_0$ is the linear-cavity Hamiltonian, $|\Phi_{n,m}\rangle$ is the Floquet eigenstate, and $\varepsilon_{n,m} = \varepsilon_n - m\Omega$ is the Floquet eigenvalue associated with the *m*-th replica of the *n*-th LG mode. Here, $\Omega = \pi/L$ denotes the Floquet "driving frequency", and $\varepsilon_n = \xi_n/(2L)$ is the eigenvalue of the *n*-th LG mode, determined by its single-pass Gouy phase $\xi_n$. In the Floquet framework, cavity degeneracy occurs when the eigenvalue of the (*n*, *m*)-th state coincides with that of the fundamental LG$_{00}$ mode (Fig. 1c), i.e. $\varepsilon_{n,m} = \varepsilon_{0,0}$, which yields the same degeneracy condition as in Eq. (2). Owing to the periodic nature of the cavity, such degeneracy with index pair ($u$, $v$) supports an infinite number of degenerate modes, with their indices ($n$, $m$) being integer multiples of ($u$, $v$).

To analyze a nonlinear MPC, we apply perturbation theory by expanding the ground-state eigenfunction in the Floquet basis: $|\Psi_{0,0}\rangle = \sum_{n,m} C_{n,m} |\Phi_{n,m}\rangle$, where the expansion coefficients $C_{n,m}$ are given by

$$C_{n,m} = \frac{-bC_{0,0}|C_{0,0}|^2 \Theta_{n,m}(d)}{\varepsilon_{0,0} - \varepsilon_{n,m}}. \tag{4}$$

The overlap integral $\Theta_{n,m}(d)$ is defined as

$$\Theta_{n,m}(d) = \frac{1}{2d_{\text{eff}}} \langle \Phi_{n,m} | S(z,d) | \Phi_{0,0} |^2 | \Phi_{0,0} \rangle, \tag{5}$$

where $S(z, d)$ is a periodic Heaviside function defining the Kerr-medium regions. Within the Floquet framework, the expansion coefficient of the LG$_{0n}$ mode is given by $c_n = \sum_m C_{n,m}$. A detailed derivation is provided in Methods.

This Floquet and perturbation model enables a detailed examination of soliton stability conditions in nonlinear MPCs. Figures 3a-f show beam-profile variations for $d/L \to 0$ at $b$=0.5 rad in degenerate and non-degenerate MPCs, while the results for $2d$=$2d_{\text{MCS}}$ are shown in Figs. 3g-i. These results correspond to the conditions A, B, and C in Figs. 2a-c.

For a nearly degenerate solid MPC with $F/L$=0.663 [close to the degenerate condition A with $F/L$=2/3 and ($u$, $v$)=(3, 2)], the radial profile of the eigenstate ($|U(r)|$) deviates from a Gaussian shape, exhibiting ripples at large radii (Fig. 3a). The spatial-mode expansion reveals significant contributions from the higher-order LG modes (Fig. 3b). To validate the perturbation approach, we numerically compute the eigenstate profiles using the Fox-Li iteration algorithm[68,71] (red lines and symbols in Figs. 3a-b;

see SM Section S3). Although some discrepancies arise when *b* is large, the perturbation approach effectively captures the emergence of higher-order modes with accurate indices. Further stability analysis (see SM Section S4) shows that multimode coupling in degenerate solid MPCs significantly disrupts stable propagation of the eigenstate (Fig. 3c), indicating that cavity degeneracy hinders soliton stabilization.

In contrast, under the non-degenerate condition B (*F/L*=0.8), the eigenstate maintains a nearly Gaussian profile (Fig. 3d), with the expansion coefficients showing negligible contributions from higher-order modes (Fig. 3e). The eigenstate exhibits stable discrete-soliton modes in the non-degenerate solid MPC (Fig. 3f).

The stark contrast between the degenerate and non-degenerate MPCs (Figs. 3a-f) is consistent with the phase diagram shown in Fig. 2a and can be explained by Eq. (4). When $d/L \to 0$, the overlap integral $\Theta_{n,m}$ is typically non-zero. Consequently, the degenerate condition ($\varepsilon_{0,0} = \varepsilon_{n,m}$) causes the expansion coefficient $C_{n,m}$ to diverge, driving substantial energy transfer from the $LG_{00}$ mode to higher-order modes. We note that, while previous studies have revealed the importance of multimode coupling in degrading beam quality in degenerate MPCs[61,62], our results provide a novel perspective based on multimode solitons, offering a comprehensive understanding of the underlying mechanisms.

## **Mechanisms for MCS Medium Length**

For the condition C, where the MPC is nearly degenerate but the medium length is $2d=2d_{MCS}$, our Floquet and perturbation model analysis shows that multimode coupling is effectively suppressed (Figs. 3g-h), and the eigenstate represents a soliton mode that propagates stably within the nonlinear MPC (Fig. 3i). This observation is consistent with Fig. 2b and can be attributed to the destructive interference among multimode wave components within the Kerr medium, which diminishes the overlap integral $\Theta_{n,m}(d)$ in Eq. (5). Accordingly, we refer this specific medium length as the mode-coupling-suppression (MCS) length. The effective suppression of higher-order modes under the MCS condition enables quasi-single-mode soliton propagation.

The Floquet and perturbation model can well capture the key features of the phase diagrams in Figs. 2a-c, including the correlation between high beam instability and the degeneracy points in the limit $d/L \to 0$, as well as the recovery of beam quality when $2d=2d_{\text{MCS}}$. Here, the beam-mode stability is characterized by the higher-order-mode contributions to the total energy: $\chi = \frac{\sum_{n \neq 0}|C_n|^2}{\sum_n |C_n|^2}$ (see SM Section S5). Although the model neglects space–time coupling, it provides an effective framework for understanding the mechanisms underlying the observed phase diagrams. We note, however, that space-time coupling *does* modify the details of the stability landscape. One particular example is that, in gas-filled MPCs with $d/L$=1, spatio-spectral inhomogeneity emerges at high nonlinearity even slightly away from the degenerate points (Fig. 2c). Such disruption can also be observed in solid MPCs, when comparing Fig. 2a and Fig. S3a. This behavior arises because nonlinear propagation introduces space-time coupling that perturbs the ideal destructive interference conditions [$\Theta_{n,m}$=0; see Eqs. (4-5)], thereby introducing multimode coupling and beam instability.

For a degenerate mode $(n, m)$, $\Theta_{n,m}(d)$ in Eq. (5) can be calculated analytically (see SM Section S6). For a given degenerate cavity with indices $(u, v)$, the multimode coupling is suppressed only when the expansion coefficients of all the degenerate modes ($n$, $m$), derived from ($u$, $v$), vanish ($C_{n,m}=0$). By setting $\Theta_{n,m}(d) = 0$, we obtain $d_{\text{MCS}}$:

$$4u \arctan\left(\frac{d_{\text{MCS}}/L}{\sqrt{2F/L-1}}\right) = 2k\pi, \ k = 1, 2, \cdots, v, \tag{6}$$

where the left-hand side represents the accumulated Gouy phase difference within the Kerr medium. Equation (6) tells that for each ($u$, $v$), there are $v$ distinct $d_{\text{MCS}}$ values that satisfy this condition. A detailed derivation is provided in SM Section S5. In Figs. 4a, we use ($u$, $v$)=(4, 3) as an example, and plot the Gouy-phase differences [Eq. (6)] and $\Theta_{n,m}(d)$, with the corresponding $d_{\text{MCS}}$ values indicated.

Figure 4a also explains why gas-filled MPCs can support high-quality beam propagation and efficient nonlinear light-matter interaction. First, the coefficients $C_{n,m}$ generally decrease oscillatory with increasing $d/L$. This behavior is analogous to phase-matching in nonlinear optics[72], where a thicker nonlinear medium leads to destructive interference among multimode wave components. Second, in degenerate cavities, $d/L=1$ always results in vanishing overlap integrals for all the degenerate modes ($n$, $m$), effectively suppressing multimode coupling. Therefore, gas-filled MPCs can be regarded as a special case of the MCS condition.

**Supercontinuum Generation in MPCs Operating at High Nonlinearity**

The discovery of the MCS condition opens the possibility for high-quality supercontinuum generation (SCG) and pulse compression in MPCs operating at high nonlinearity. To evaluate this potential, we conduct full space-time-coupled NLSE simulations. The MPC is configured under the degenerate condition, with geometry parameters summarized in Table 1. Fused silica is selected as the Kerr medium, with realistic material parameters implemented (Table 1). $\chi_K$, $\tau_1$, and $\tau_2$ are the Raman response coefficients (see Methods). The medium length is set to $2d=2d_{MCS}$=8.47 cm, corresponding to $d_{MCS} \approx L/3$. Notably, because of the relatively thick medium length, material GDD must be accounted for. We therefore introduce a compensating negative GDD of -1560 fs$^2$ per bounce on the cavity mirrors, following the approaches used in recent MPC studies[23,46]. The input laser pulses have a duration of $\tau_p$=170 fs and pulse energy of $E_0$=1.64 μJ at $\lambda_0$=1030 nm, corresponding to a peak intensity $I_0$ of $4.8\times10^{10}$ W/cm$^2$ and a SNLP of $1.5\pi$.

Figure 4b shows the evolution of transverse beam profiles over 18 passes (9 roundtrips), demonstrating stable propagation. Because the overlap integral Θ is small near $2d=2d_{MCS}$, the beam quality remains robust against small variations in medium length around $2d_{MCS}$. For comparison, replacing the thick Kerr medium with a thin fused-silica plate ($d$=1.0 mm), as commonly employed in solid MPCs, leading to beam collapse and severe beam-quality degradation after only 6 roundtrips (Fig. 4c). The spectrum bandwidth obtained under the MCS condition (Fig. 4b) is sufficient for direct pulse compression from 170 fs to ~12.3 fs (Fig. 4d), achieving >13-fold compression in a single-stage, all-solid MPC compressor with only 9 roundtrips. Importantly, the spatio-spectral homogeneity remains as high as 0.93, even under such strong nonlinearity (Fig. 4e), confirming the effectiveness of the MCS condition in preserving beam quality.

To further examine the robustness of the MCS condition, we further simulate three perturbations: (i) displacing the Kerr medium by 1 cm from the cavity center (~4% of the total cavity length) to mimic misalignment, (ii) varying the medium length to $2d=1.9d_{MCS}$ and $2.1d_{MCS}$ to mimic fabrication inaccuracy, and (iii) altering the group velocity dispersion of the Kerr medium by ±5% to simulate uncertainties in material parameters. In all the cases, stable beam propagation and high-quality SCG can be preserved (see SM section S7), demonstrating that the MCS condition is robust against realistic experimental imperfections.

**<u>Discussion and Conclusion</u>**

In addition to multimode coupling, the strength of nonlinearity in MPCs is constrained by the critical power of the Kerr medium ($P_{cr}$). To prevent catastrophic beam collapse and material damage in a long Kerr medium, the condition $P_0/P_{cr}<1$ must be satisfied, which yields a maximum SNLP:

$$b_{\max} = 2\pi \frac{k}{u}, \quad k = 1, 2, \cdots, v. \tag{7}$$

Equation (7) is in agreement with previous studies on gas-filled MPCs[63], which predict a maximum SNLP of $2\pi$ with $k=u$.

In this study, we extend this analysis to MPCs with variable medium lengths. In such cases, Eq. (7) no longer provides an accurate constraint. Instead, the condition must be modified by requiring the self-focusing length $z_{SF}$ to exceed the medium length $2d$, ensuring that the focal point remains outside the medium. This leads to a modified maximum SNLP:

$$b_{\max} = 2\pi \frac{k}{u} \left[ \sqrt{\left(\frac{0.367}{2\tan\left(\frac{k\pi}{2u}\right)}\right)^2 + 0.0219} + 0.852 \right]^2. \tag{8}$$

A detailed derivation is provided in Methods. In SM Fig. S7, we plot $b_{max}$ as a function of $k/u$. Notably, Eq. (8) predicts higher $b_{max}$ values for thin Kerr media, while it smoothly approaches Eq. (7) in the limit $k/u \to 1.0$.

Equation (8), combined with our multimode-coupling analysis, provides valuable guidelines for designing nonlinear MPCs at high nonlinearity. Thin Kerr media can, in principle, support higher peak powers and push $b_{max}$ beyond $3\pi$ (see SM Fig. S7), but strong multimode coupling in this regime undermines soliton stability and prevents stable beam propagation. Gas-filled MPCs operate close to the $k/u \to 1.0$ limit, where multimode coupling is suppressed, yet the critical power condition restricts the maximum SNLP to $2\pi$. Our study identifies an intermediate regime: by operating under the MCS condition, multimode coupling can be effectively suppressed, while a relatively short medium length allows high input peak power. This strategy is particularly relevant for solid-state MPCs, enabling stable operation at high nonlinearity. In the example presented in Figs. 4b-e, we achieve an SNLP of ~$1.5\pi$, which approaches the theoretical maximum predicted for $k/u = 0.545$.

Although our analysis underscores the superior performance of gas-filled MPCs, solid MPCs offer unique flexibility through the tailored design of Kerr media, as we understand the mechanism to suppress multimode coupling in nonlinear MPCs. While our study focuses on a single bulk Kerr medium placed at the cavity center, the same MCS mechanism can be extended to configurations involving multiple periodic or nonperiodic, centro-symmetric, or asymmetric distributions of Kerr media. Various MPC geometries beyond the Herriott-type configuration can also be explored. These advancements open up new possibilities for applications of optical cavities in supercontinuum generation and other nonlinear processes.

In conclusion, we have investigated the stability of multimode solitons in nonlinear MPCs with strong Kerr nonlinearity. Using Floquet analysis and first-order perturbation theory, we elucidated the critical role of multimode coupling in soliton destabilization. Importantly, we identified the novel MCS condition that enables soliton stabilization in degenerate MPCs with a nonlinear phase of up to $1.5\pi$ per pass in solid MPCs, significantly exceeding the current state-of-the-art. Our findings provide a unified framework that explains the disparity between gas-filled and solid MPCs, offering new opportunities for advanced nonlinear MPCs with tailored Kerr media for high-power and high-efficiency ultrafast applications.

**Table 1. Key parameters for the NLSE simulations.**

| | |
|---|---|
| Cavity Geometry | |
| Cavity length, $2L$ (cm) | 25 |
| Focal length, $F$ (cm) | 6.79 |
| Medium length, $2d$ (cm) | 8.47 |
| GDD at each cavity mirror (fs$^2$) | -1560 |
| Kerr Medium (Fused Silica) | |
| $n_0$ | 1.45 [73] |
| $n_2$ (m$^2$W$^{-1}$) | 2.4×10$^{-20}$ [74] |
| Group velocity dispersion | Calculated using data in Ref. 73 |
| $\chi_K$ | 0.2 [65] |
| $\tau_1$ (fs) | 20 [65] |
| $\tau_2$ (fs) | 40 [65] |
| Laser Parameters | |
| $\tau_p$ (fs) | 170 |
| $\lambda_0$ (nm) | 1030 |
| $E_0$ (μJ) | 1.64 |
| $I_0$ (W/cm$^2$) | 4.8×10$^{10}$ |

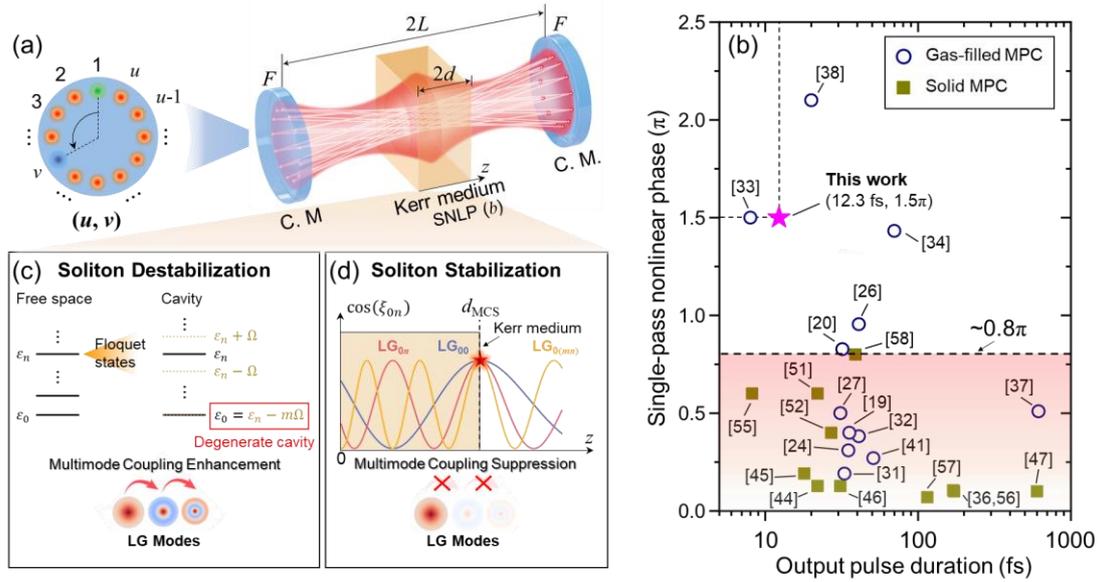

**Figure 1. Mechanisms of stabilization and destabilization of multimode solitons in nonlinear MPCs. (a)** Schematic of a nonlinear MPC geometry. C.M.: Concave mirror. **Inset:** Laser-spot distribution on a Herriott-type MPC with a degeneracy defined by indices $(u, v)$. The green spot marks the initial position where the incident laser beam strikes, while the blue spot indicates its position after the first roundtrip. **(b)** Summary of state-of-the-art nonlinear MPCs for supercontinuum generation and pulse compression. The dashed line indicates the SNLP limit of existing solid MPCs. **(c-d)** Mechanisms underlying the destabilization and stabilization of multimode solitons in nonlinear MPCs.

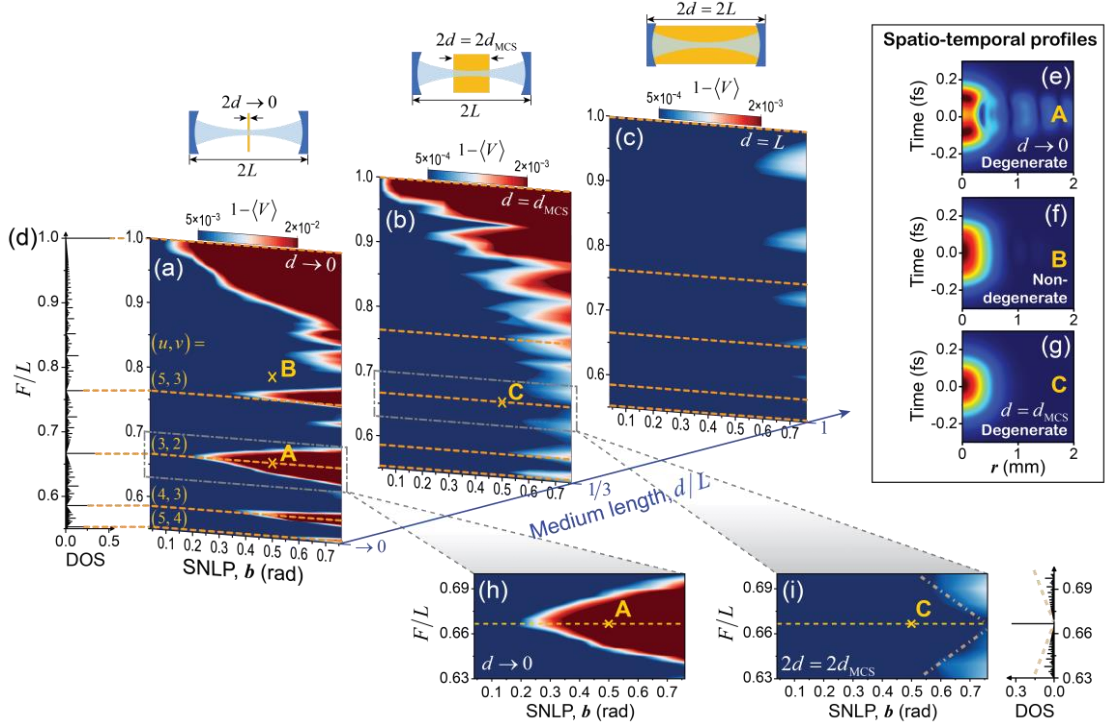

**Figure 2. Stabilization landscape of nonlinear MPCs. (a-c)** Phase diagrams of output beam inhomogeneity as a function of cavity geometry $F/L$ and SNLP $b$, for medium length corresponding to **(a)** a thin plate ($2d \to 0$), **(b)** the MCS length ($2d=2d_{MCS}$), and **(c)** a gas-filled MPC ($2d=2L$). Conditions of A, B, and C are labeled. Dashed lines indicate degeneracy geometries with $(u, v)$=(5, 3), (3, 2), (4, 3), and (5, 4). **(d)** DOS as a function of $F/L$, with the DOS peaks corresponding to the degeneracy geometries (dashed lines). **(e-g)** Spatio-temporal profiles of the output beams after 20 roundtrips for Condition A, B, and C, as labeled in **(a-c)**. **(h-i)** Zoomed-in views of the phase diagrams near the degeneracy point $(u, v)$=(3, 2), as illustrated by the dashed-dotted boxes in **(a-b)**. DOS results for the same $F/L$ region are shown in the right panel.

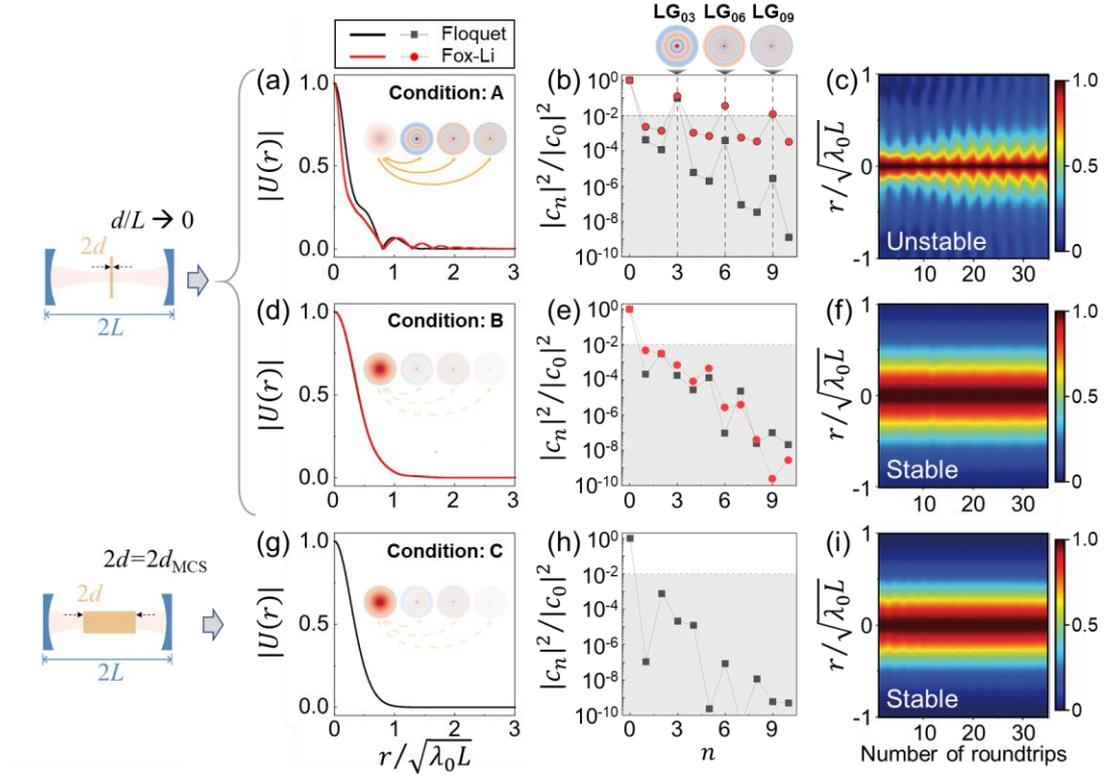

**Figure 3. Stabilization and destabilization of multimode solitons. (a)** Radial profiles of the eigenstate for a nearly degenerate solid MPC with $F/L=0.663$ (close to Condition A) with SNLP $b=0.5$, obtained from Floquet and perturbation analysis (black line) and the Fox-Li algorithm (red line). **Inset:** Illustration of energy transfer from the $LG_{00}$ mode to higher-order LG modes. **(b)** Normalized expansion coefficients $|c_n|^2/|c_0|^2$ of the eigenstates shown in **(a)**. The horizontal dashed line indicates the amplitude of $|c_n|^2$ being 1% of the amplitude of the fundamental mode $|c_0|^2$. The LG modes of $LG_{03}$, $LG_{06}$, and $LG_{09}$ are labeled. **(c)** Stabilization analysis of the eigenstate shown in **(a)**. **(d-f)** Same as **(a-b)**, but for the results at Condition B. **(g-h)** Same as **(a-b)**, but for the results near Condition C.

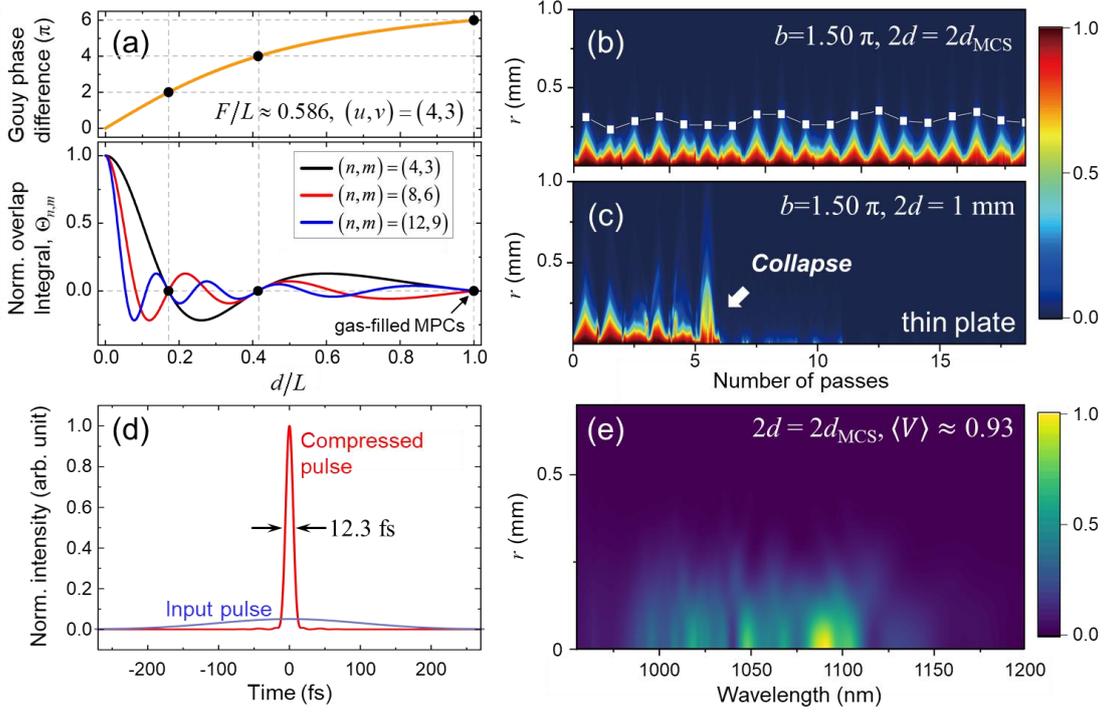

**Figure 4. MCS mechanism stabilizing soliton propagation. (a)** Gouy phase difference and normalized overlap integral $\Theta_{n,m}$ as a function of $d/L$ for $F/L$=0.586 and $(u, v)$=(4, 3). Dashed lines indicate the medium lengths where the Gouy phase accumulates by integer multiples of $2\pi$, and $\Theta_{n,m}$=0 for all degenerate modes. **(b)** Simulated propagation of femtosecond laser pulses within a degenerate cavity with $(u, v)$=(11, 9) under SNLP $b$=1.5$\pi$ and medium length $2d$=$2d_{MCS}$. Beam radii on a cavity mirror are indicated by symbols. **(c)** Same simulation conditions and cavity geometry as **(b)**, but for a thin-plate Kerr medium ($2d$=1 mm). **(d)** Temporal profile of the compressed output pulse (red line) corresponding to the spectrum obtained from the simulation shown in **(b)**. The blue line indicates the temporal profile of the input pulse. **(e)** Radial distribution of the output pulse spectrum obtained from the simulation in **(b)**.

**Methods**

**NLSE simulations.** Nonlinear MPCs are simulated using the equivalent-lens sequence model, where cavity mirrors are replaced by thin focusing lenses and the beam is assumed to propagate in the forward direction. The Kerr medium, with a thickness of 2$d$, is placed at the cavity center. The forward-propagation NLSE with radial symmetry is given by[65]

$$\frac{\partial U}{\partial z} = \frac{i}{2n_0 k_0} \mathcal{T}^{-1} \nabla_\perp^2 U + i\mathcal{D}U + i\frac{\omega_0}{c} n_2 \mathcal{T} \left[ (1-\chi_K)|U|^2 + \chi_K \int_{-\infty}^{t} h(t-t')|U(t')|^2 dt' \right] U, \quad (9)$$

where $U$ is the complex field amplitude, $t$ is the retarded time $t - z/v_g$, with $v_g$ being the group velocity near the carrier frequency $\omega_0$, and $k_0$ is the vacuum wavevector. The dispersion operator is $\mathcal{D} = \frac{k''}{2}(i\partial_t)^2$. The operator $\mathcal{T} = \left(1 + \frac{i\partial_t}{\omega_0}\right)$ accounts for self-steepening. The Raman response is parameterized by $\chi_K$ and the response function

$$h(t) = \frac{2}{3}\frac{\tau_1^2 + \tau_2^2}{\tau_1 \tau_2^2} e^{-t/\tau_2} \sin(t/\tau_1), \quad (10)$$

with $\tau_1$ and $\tau_2$ being Raman time constant. The effect of the cavity mirrors is incorporated through the thin-lens transformation: $U' = U e^{-i\pi r^2/\lambda_0 F}$.

The NLSE is numerically solved using the split-step Fourier method[75]. To account for large variations in beam radius during propagation, we implement a non-uniform radial grid defined by the transformation $r = r_0(e^y - 1)$, with $y$ uniformly discretized. Here, $r_0$ is chosen to match the beam waist radius of the corresponding cavity eigenmode. Numerical accuracy is controlled by maintaining a local error of O($dz^3$), where $dz$ is the propagation step size.

**Spatio-spectral inhomogeneity.** Quantitatively, the spatio-spectral homogeneity can be characterized by the spectral overlap integral[57] $V(r) = \frac{\left\{\int_\lambda [I(\lambda,r)I(\lambda,0)]^{1/2} d\lambda\right\}^2}{\int_\lambda I(\lambda,r)d\lambda \cdot \int_\lambda I(\lambda,0)d\lambda}$, where $I(\lambda,r)$ represents the spectral intensity of the output beam at radial coordinate $r$. The average overlap integral across the output beam is given by $\langle V \rangle = \frac{\int V(r)I(r)rdr}{\int I(r)rdr}$.

**Floquet theory.** We begin with the simplified NLSE is given by

$$i\frac{\partial U}{\partial z} = -\frac{1}{2n_0 k_0}\nabla_\perp^2 U + V_l(r,z)U + bV_k(r,z,U)U, \tag{11}$$

where the cavity-mirror potential $V_l(r,z)$ is given by

$$V_l(r,z) = \frac{\pi r^2}{\lambda_0 F}\sum_n \delta(z - (2n+1)L), \tag{12}$$

and the Kerr nonlinear term $V_k(r,z,U)$ is

$$V_k(r,z,U) = -n_2 k_0 I_0(z=0)S(z,d)|U|^2 \equiv -\frac{1}{2d_{\text{eff}}}S(z,d)|U|^2. \tag{13}$$

Here, $V_l(r,z)$ represents the cavity-mirror effect modeled using the thin-lens approximation. $V_k(r,z,U)$ describes the self-focusing effect induced by Kerr nonlinearity. $S(z,d)$ is a periodic Heaviside function defining the medium length:

$$S(z,d) \equiv \begin{cases} 1, & |z| \leq d \\ 0, & \text{else} \end{cases}, \quad z \in [-L, L], \text{ and } S(z+2L) = S(z). \tag{14}$$

For convenience, we define an effective length $2d_{\text{eff}} = 2z_0 \arctan(d/z_0)$, where $z_0$ is the Rayleigh distance[63]. The nonlinear phase per pass is then given by

$$b = n_2 k_0 I_0 \cdot 2d_{\text{eff}}. \tag{15}$$

We use Floquet theory to analyze the linear contribution in Eq. (8). The linear Hamiltonian is given by $H_0(r,z) = -\frac{1}{2n_0 k_0}\nabla_\perp^2 + V_l(r,z)$, and the corresponding Floquet eigenequation is:

$$\left[H_0(r,z) - i\frac{\partial}{\partial z}\right]|\Phi_n(r,z)\rangle = \varepsilon_n|\Phi_n(r,z)\rangle. \tag{16}$$

Since the Floquet Hamiltonian is periodic along $z$, its eigenstate $|\Phi_n(r,z)\rangle$ has an infinite number of replicas, $|\Phi_{n,m}(r,z)\rangle$, with Floquet eigenvalue $\varepsilon_{n,m} = \varepsilon_n - m\Omega$, where $\Omega = \pi/L$ is the "driving frequency".

In the subspace with zero angular momentum, the Floquet state is expressed as

$$|\Phi_{n,m}(r,z)\rangle = \frac{\sqrt{2/\pi}}{w(z)} L_n\left[2\frac{r^2}{w^2(z)}\right] e^{-\frac{r^2}{w^2(z)}} e^{-ik\frac{r^2}{2R(z)}} e^{-i\frac{1}{2}\xi_n(z)+i\frac{\xi_n(L)}{2L}z-im\Omega z} \equiv$$

$$\psi_n(r,z) e^{i\frac{\xi_n(L)}{2L}z-im\Omega z}, \quad (17)$$

where $\psi_n(r,z) = \frac{\sqrt{2/\pi}}{w(z)} L_n\left[2\frac{r^2}{w^2(z)}\right] e^{-\frac{r^2}{w^2(z)}} e^{-ik\frac{r^2}{2R(z)}} e^{-i\frac{1}{2}\xi_n(z)}$ is the linear-cavity eigenstate, also known as the LG$_{0n}$ mode, $L_n$ represents the $n$-th order Laguerre polynomial, $\xi_n(z)$ is the Gouy phase of the LG$_{0n}$ mode:

$$\xi_n(z) = 2(2n+1)\arctan\left(\frac{z/L}{\sqrt{2F/L-1}}\right). \quad (18)$$

The accumulated Gouy phase upon one pass through the cavity is $\xi_n(L)$. The Floquet eigenvalue, thus, is given by:

$$\varepsilon_{n,m} = \frac{\xi_n(L)}{2L} - m\Omega. \quad (19)$$

Cavity degeneracy occurs when the eigenvalue of $(n, m)$-th state coincides with that of the ground state, $\varepsilon_{n,m} = \varepsilon_{0,0}$, which yields the same degeneracy condition as shown in Eq. (2).

**Perturbation theory.** We apply perturbation theory to derive the ground-state solitons in nonlinear MPCs. Since the soliton state shares the same periodicity as the cavity, the ground-state soliton $|\Psi_{0,0}(r,z)\rangle$ and its eigenenergy $\epsilon_{0,0}$ can be expanded as a superposition of the Floquet eigenstates and eigenvalues:

$$|\Psi_{0,0}(r,z)\rangle = C_{0,0}|\Phi_{0,0}(r,z)\rangle + b\sum_{n'\neq 0,m'\neq 0}\tilde{C}_{n',m'}|\Phi_{n',m'}(r,z)\rangle, \quad (20)$$

$$\text{and } \epsilon_{0,0} = \varepsilon_{0,0} + b\Delta_{0,0}. \quad (21)$$

Here, we introduce the perturbation terms by assuming $b \ll 1$. The nonlinear eigenequation is

$$\left[H(r,z) - i\frac{\partial}{\partial z}\right]|\Psi_{0,0}(r,z)\rangle = \epsilon_{0,0}|\Psi_{0,0}(r,z)\rangle, \quad (22)$$

where the nonlinear Hamiltonian is $H(r,z) = H_0(r,z) + bV_k(r,z,U)$. By substituting Eqs. (17-18) into Eq. (19), we obtain

$$\sum_{n'\neq 0, m'\neq 0}(\varepsilon_{0,0} - \varepsilon_{n',m'})b\tilde{C}_{n',m'}|\Phi_{n',m'}\rangle + b\Delta_{0,0}C_{0,0}|\Phi_{0,0}\rangle = V_k C_{0,0}|\Phi_{0,0}\rangle. \quad (23)$$

By taking the inner product with $\langle\Phi_{n,m}|$ and using the orthogonal relation of the Fluquet eigenstates, we arrive

$$\tilde{C}_{n,m} = \frac{C_{0,0}\langle\Phi_{n,m}|V_k|\Phi_{0,0}\rangle}{\varepsilon_{0,0}-\varepsilon_{n,m}}. \quad (24)$$

We further evaluate the overlap integral:

$$\langle\Phi_{n,m}|V_k|\Phi_{0,0}\rangle = -\frac{1}{2d_{\text{eff}}}|C_{0,0}|^2 \langle\Phi_{n,m}|S(z,d)|\Phi_{0,0}|^2|\Phi_{0,0}\rangle \equiv -|C_{0,0}|^2\Theta_{n,m}(d), \quad (25)$$

where

$$\Theta_{n,m}(d) = \frac{1}{2d_{\text{eff}}}\langle\Phi_{n,m}|S(z,d)|\Phi_{0,0}|^2|\Phi_{0,0}\rangle. \quad (26)$$

Numerically, $\Theta_{n,m}(d)$ can be calculated by

$$\Theta_{n,m}(d) = \frac{1}{2d_{\text{eff}}}\frac{\pi}{L}\int_{-d}^{d}\left(\int_0^{+\infty} \phi_n^*(r,z)|\phi_0(r,z)|^2\phi_0(r,z)e^{i2n\left[\frac{1}{2}\xi_0(z)-\frac{\xi_0(L)}{2L}z\right]+im\Omega z}rdr\right)dz, \quad (27)$$

where $\phi_n(r,z) = \frac{\sqrt{2/\pi}}{w(z)}L_n\left[2\frac{r^2}{w^2(z)}\right]e^{-\frac{r^2}{w^2(z)}}e^{-ik\frac{r^2}{2R(z)}}$. Equation (25) is an alias for Eq. (5) in the main text.

**Cavity degeneracy and density of states.** In the Floquet framework, the normalized DOS is defined as[70]:

$$D(\varepsilon) = \frac{1}{N}\sum_{n,m}\delta(\varepsilon - \varepsilon_{n,m}), \quad (28)$$

where $N$ is the total number of states. All distinct Floquet states can be indexed by eigenvalues within the first Floquet Brillouin zone (FBZ) $[\varepsilon_0, \varepsilon_0 + \Omega]$, For a degenerate cavity, the DOS peaks at the degenerate eigenvalues, whereas for a non-degenerate cavity, it approaches zero as $N \to \infty$. Thus, for a cavity characterized by degenerate indices $(u, v)$, the DOS can be explicitly expressed as

$$D = \begin{cases} \frac{1}{u}, & \text{for a } (u,v) \text{ degenerate cavity} \\ 0, & \text{for a non} - \text{degenerate cavity} \end{cases}. \qquad (29)$$

**Critical power constraint.** The critical power for self-focusing in a Kerr medium is approximately[76]

$$P_{\text{cr}} \approx \lambda_0^2/(2\pi n_0 n_2). \qquad (30)$$

According to Eq. (15), the incident peak power is related to SNLP as

$$P_0 = \frac{\pi w_0^2 \lambda_0 b}{8\pi n_0 n_2 d_{\text{eff}}}, \qquad (31)$$

where $w_0$ is the beam waist.

For gas-filled MPCs, imposing the critical-power condition $P_0/P_{\text{cr}} < 1$ yields

$$b < \frac{4 d_{\text{eff}}}{z_0}, \qquad (32)$$

where the effective medium length is defined as $d_{\text{eff}} = z_0 \arctan\left(\frac{d/L}{\sqrt{2F/L-1}}\right)$. This leads to the maximum SNLP:

$$b_{\max} = 4 \arctan\left(\frac{d/L}{\sqrt{2F/L-1}}\right). \qquad (33)$$

Finally, under the MCS condition, substituting Eq. (6) into Eq. (33) gives Eq. (7) in the main text.

We also consider the relaxed constraint, where $P_0/P_{\text{cr}} > 1$ is allowed, while the self-focusing point remains outside the Kerr medium ($z_{\text{SF}} > 2d$). The self-focusing length can be empirically estimated as [77]

$$z_{\text{SF}} = \frac{0.367 z_0}{\sqrt{\left[(P_0/P_{\text{cr}})^{\frac{1}{2}} - 0.852\right]^2 - 0.0219}}, \qquad (34)$$

where $z_0$ is the Rayleigh distance. This yields

$$b < \frac{4 d_{\text{eff}}}{z_0}\left[\sqrt{\left(0.367 \frac{z_0}{2d}\right)^2 + 0.0219} + 0.852\right]^2. \qquad (35)$$

Substituting Eq. (6) into Eq. (35) leads to Eq. (8) in the main text. In SM Fig. S7, we plot $b_{max}$ as a function $k/u$ obtained from Eqs. (7) and (8).

**Data availability**

The minimum dataset required to reproduce the conclusions is provided in the main text and supplementary information. Additional data can be obtained upon request from the corresponding author, Z. T.